\newcommand{\bea}{\begin{eqnarray}}
\newcommand{\eea}{\end{eqnarray}}
\newcommand{\nn}{\nonumber}
\newcommand{\Ga}{\Gamma}

\renewcommand{\a}{\alpha}

\newcommand{\s}{\sigma}

\renewcommand{\d}{\delta}

\newcommand{\bt}[1]{{\bar t}}

\newcommand{\sgn}{\mathrm{sgn}}
\documentclass[twocolumn,showpacs,preprintnumbers,amsmath,amssymb]{revtex4}
\usepackage{graphicx}% Include figure files
\usepackage{dcolumn}% Align table columns on decimal point
\usepackage{bm}% bold math
\begin{document}

\preprint{} %\hspace{20pt}
\title{Ground State Instabilities and Entanglement in the Spin-Boson Model}

\author{Ru-Fen Liu}
 \email{fmliu@phys.ncku.edu.tw}%Lines break automatically or can be forced with \\
\author{Chia-Chu Chen}%
 \email{chiachu@phys.ncku.edu.tw}
\affiliation{%
National Cheng-Kung University, Physics Department, 70101, 1
University Road, Tainan, Taiwan, R. O. C. }%

\begin{abstract}
Ground state instabilities of the spin-boson model is studied in
this work. The existence of sequential ground state instabilities
is shown analytically for arbitrary detuning in the two-spin
system. In this model, extra discontinuities of
concurrence(entanglement measure) are found in the finite system,
which do not appear in the on-resonant model. The above results
remain intact by including extra boson modes. Moreover, by
including extra modes, it is found that ground state entanglement
can be obtained and enhanced even in the weak coupling regime.
\end{abstract}

\maketitle

Entanglement\cite{EPR} has been recognized as the essential
element of quantum information science. This is due to the fact
that the nonlocal quantum coherent nature of entanglement can be
used as a resource for implementing quantum information
protocols\cite{Issac}. Recently, the concept of entanglement has
also been introduced to the investigation of quantum phase
transition(QPT)\cite{scaling} which is induced by quantum
fluctuations and therefore can occur even at zero temperature.
More precisely, QPT can be identified as the appearance of
non-analyticity in the ground state energy. For the case of
spin-boson model treated in this work, a phenomenon closes to QPT
known as Ground State Instabilities(GSI) also occurs in finite
system. One will see that non-analyticity arises due to level
crossing\cite{QPT} which indicates the instability of ground
state. Similar problems of QPT in the weak coupling and
thermodynamical limit have been studied by many
authors\cite{Hepp}. In addition, the related problem of
entanglement in the so-called Dicke model(DM)\cite{Dicke} has also
attracted much attention\cite{Milburn} recently. One of the
interesting results of the Dicke model is the sequence of GSI in
arbitrary finite-atom system\cite{Buzek} which has been overlooked
in the thermodynamical limit. More interestingly, at these
infinite sequential instabilities, Bu\v{z}ek et al. show that
there are corresponding discontinuities appearing in the
ground-state entanglement of the reduced atomic system. However,
it has been pointed out by K. Rza\.{z}ewski and K.
W\'{o}dkiewicz\cite{Wod} that in \cite{Buzek} gauge invariance is
spoiled by not including the $A^2$ term of the minimal coupling
hamiltonian. Furthermore, they also pointed out that without the
$A^2$ contribution the hamiltonian is unbounded from below as the
coupling goes to infinity. It is obvious that any two-level atomic
system is isomorphic to a spin-1/2 system. Therefore the Dicke
model can be identified as a spin-boson interacting system.
Certainly for the spin-boson system, there is no requirement of
gauge invariance. Furthermore, by keeping finite coupling it seems
that the unbounded problem can be avoided. However, by requiring
finite coupling, one can only allow the investigation of finite
number of ground state transitions instead of the infinite
transitions in \cite{Buzek}. Even though with such restriction,
the correlation between entanglement and GIS can still be
addressed. In order to understand the relation between ground
state instability and concurrence, exact analytical results are
needed for gaining insight. Here, we discuss the generic
spin-boson model by introducing a parameter $r$ which is the
detuning of the boson mode frequency. Moreover, for more realistic
consideration, we also study the ground state instabilities in the
multi-mode model. In this work, we rigorously show the existence
of sequential ground state instabilities for arbitrary detuning in
the two-spin system. In contrast to the resonant case where GIS
and concurrence are strongly correlated, such detuning effect
leads to the disconnection of ground state instability and
concurrence. By including extra modes and tuning the frequency, it
is found that the ground state can become entangled even in the
\textit{weak} coupling regime and the entanglement is enhanced
comparing with the mono-mode on-resonant spin-boson model(SBM,
from now on, the term SBM denotes the mono-mode on-resonant
model). The plan of the paper starts by introducing the
single-mode spin-boson model, and the exact spectrum is then
presented for two-spin system. In the next section we will show
that ground state instability is a generic phenomenon of
spin-boson model. Section III provides the analysis of ground
state entanglement by calculating the concurrence. In this
section, we establish the fact that ground state instability is
not directly correlated with the analyticity of concurrence. The
study of ground state instabilities of the two-mode model is
presented in section IV. By including extra modes, for two spins,
it is found that there exists a region of detuning where
\textit{enhanced }ground state entanglement can be obtained even
in the \textit{weak coupling} regime. Finally, a brief summary is
given in the last section.

\section{The Single-Mode Model and Its Spectrum} To begin with, we discuss the
general method to solve the $N$-spin model. The system is $N$
spins interacting with a mono-mode boson field. The Hamiltonian of
the total system in the interaction hamiltonian is given by
($\hbar=1$) \bea H &=& \omega_0J_z + \omega a^{\dagger}a+
gJ_+a+g^*J_-a^{\dagger}\eea where $J_{\a}\equiv
\frac{1}{2}\sum_{j=1}^N\s_j^{\a}$, $\a$ can either be $\{+, -\}$
for raising and lowering operations or $\{x, y, z\}$,
$a(a^{\dagger})$ is the boson annihilation(creation) operator.
${\s^{\a}}$ are the Pauli matrices. $\omega_0$ is the level
spacing of the spin and $\omega$ indicates the frequency of the
boson mode. We have assumed that these spins couple to the boson
mode with the same strength $g$. This detuned spin-boson
model(DSBM) can in principle be solved exactly\cite{Cumming}. In
this work, we extend the method of
Swain\cite{swain} to diagonalize DSBM. The Hamiltonian can be separated by $H=H_0+H_I$: \bea H_0 &=& J_z+a^{\dagger}a \\
H_I &=& r a^{\dagger}a+ \kappa J_+a+\kappa^*J_-a^{\dagger} \eea
where $r\equiv \omega/\omega_0-1$ and $\kappa\equiv g/\omega_0$.
The parameter $r$($-1<r<\infty$) is related to the detuning which
is usually defined by $\omega-\omega_0$ in quantum optics. $H_0$
is the so-called excitation operator\cite{Hepp}. To obtain the
spectrum, one uses the fact that $\{H, H_0, H_I, J^2\}$ form a
maximally compatible set, where $J^2=J^2_x+ J^2_y+J^2_z$. For a
$N$-spin system, we focus on $j=N/2$ which is relevant to the
ground state discussions. Due to the commutative $J^2$, the matrix
of $H$ is automatically block diagonal by each $j$ in the basis of
$H_0$ which is denoted by $\{|j,m\rangle_A|n\rangle_p\}$.
$|j,m\rangle_A$ and $|n\rangle_p$ are the spin states and photon
number states respectively. $\{\lambda=m+n\}$ are the eigenvalues
of $\{|j,m\rangle_A|n\rangle_p\}$. It is noted that by excluding
$ra^{\dagger}a$ our $H_0=J_z +a^{\dagger}a$ is a parameter free
operator and so does its eigenvalues. This approach helps to ease
the counting of degeneracy of $H_0$. Hence, for fixed $\lambda$,
there exists degenerate subspace such that the diagonalization of
$H$ reduces to diagonalize finite matrices of $H_I$. The
eigenstates of $H$ are denoted by $|j,\lambda,h\rangle$. Then
\begin{subequations} \bea
H|j,\lambda,h\rangle &=& E_{\lambda h}|j,\lambda,h\rangle \\
H_I|j,\lambda,h\rangle &=& h|j,\lambda,h\rangle\\
|j,\lambda,h\rangle &=& \sum_{i=\lambda-j }^{\lambda+j}
A_i^{(j,\lambda,h)} |j,\lambda-i\rangle_A|i\rangle_p \\
E_{\lambda h}&=&\lambda+h.  \eea
\end{subequations} Note that, with arbitrary
detuning, the energy eigenvalues depend not only on $\kappa$, but
also on $r$ and one should expect some new results due to these
parameters dependence. The detail form of $H_I$ for arbitrary $N$
are given in Appendix A. Since the dimension of $H_I$ becomes
bigger as the excitation number increases, most spectrum can only
be obtained numerically for $N\geq 3$. However, for two-spin, the
\textit{full} spectrum with $j=1$ can be obtained(we further
neglect the index $h$ since only the eigenvalue of $H_I$ which is
a decreasing function of $\kappa$ is needed for fixed $\lambda$.):
\begin{subequations} \bea E_{\bar{1}}&=&-1
\\ E_{0} &=& \frac{1}{2} \{r-\sqrt{8|\kappa|^2+r^2}\} \\ E_{\lambda}
&=& \lambda+\lambda
r-\frac{2}{3}\sqrt{3\a_{\lambda}}\cos{\{\frac{\pi}{3}
-\frac{\varphi_{\lambda}}{3}\}} \\ && \a_{\lambda}\equiv
(4\lambda+2)|\kappa|^2+r^2 \nn \\ && \varphi_{\lambda}\equiv
\cos^{-1}{\{\frac{3\sqrt{3}\kappa^2 r}{\sqrt{\a_{\lambda}^3}}\}}
\nn \eea \end{subequations} where $\lambda$ runs from $1$ to
infinity.

\section{Sequential Ground State Instabilities in DSBM}
Usually, the eigenenergies of a quantum system are analytic
functions of the coupling constant $\kappa$. However, there is a
possibility that when $H(\kappa)=H_0+\kappa H_I$ and $[H_0,H_I]=0$
such that $H_0$ and $H_I$ can be simultaneously diagonalized and
therefore the eingenfunctions are independent of $\kappa$ even
though the eigenvalues vary linearly with $\kappa$\cite{QPT}. As a
result, when one of the excited state is crossing with the ground
state at some critical value $\kappa=\tilde{\kappa}$,
non-analyticity appears in the ground state energy. Such
level-crossing phenomenon is called ground state instability(GSI)
which also happens in the system considered in this work. The
level crossing of SBM can be illustrated easily from the
eigenstates with $0$ and $1$ excitations. From Eq.(A3), it is
clear that as $\kappa <\sqrt{(1+r)/N}$ the energy
$E_{\bar{\frac{N}{2}}}$ is less than $E_{\bar{\frac{N}{2}}+1}$, so
the ground state is the one with zero excitation. However, for
$\kappa > \sqrt{(1+r)/N}$, the ground state is replaced by the one
with $1$ excitation since
$E_{\bar{\frac{N}{2}}+1}<E_{\bar{\frac{N}{2}}}$. At the critical
value $\kappa_{\bar{\frac{N}{2}}}=\sqrt{(1+r)/N}$, the excitation
number changes discontinuously from $\lambda=-\frac{N}{2}$ to
$\lambda=-\frac{N}{2}+1$. Obviously, when $r=0$,
$\kappa_{\bar{\frac{N}{2}}}$ reduces to the on-resonant result in
Ref.\cite{Buzek}. Buzek et al have shown numerically that the
ground state energy is non-analytic and the level crossing occurs
in sequence: $\{E_{\bar{1}}\rightarrow E_0\rightarrow
E_1\rightarrow E_2\rightarrow\ldots\}$. Due to the fact that the
spectrum for two-spin can be obtained in closed form, we can
provide an analytic proof for these sequential ground state
transitions for all $r$ if the following conditions are satisfied:
($\kappa \geq 0$, $-1<r<\infty$ and $\lambda\geq -1$): $(i)$
$\{E_\lambda\}$ are monotonic decreasing functions, except for
$\lambda=-1$. $(ii)$ For all $\lambda$,
$f(\kappa,r,\lambda)=E_{\lambda+1}-E_{\lambda}$,
$f(\kappa,r,\lambda)$ is a monotonic decreasing function with
opposite signs at small and large $\kappa$. With
$\tilde{\kappa}_\lambda$ denoted the value of level crossing which
is determined by $E_{\lambda+1}=E_\lambda$, we have $(iii)$
$\{E_{\lambda+2}>E_{\lambda+1}\}|_{
\kappa=\tilde{\kappa}_\lambda}$. Due to the absolute square of
$\kappa$ in Eq.(5), one may choose $\kappa\geq 0$ without losing
generality. The first condition guarantees the eigenenergies of
different $\lambda$ involved in the ground state level crossing at
different coupling strength. The second condition ensures that
there is \textit{only one} crossing between $E_\lambda$ and
$E_{\lambda+1}$. If $E_{\lambda+2}|_{\tilde{\kappa}_\lambda}$ is
larger than $E_{\lambda+1}|_{\tilde{\kappa}_\lambda}$, then the
crossing $\tilde{\kappa}_{\lambda+1}$ which is determined by the
equation $E_{\lambda+1}=E_{\lambda+2}$ must be larger than
$\tilde{\kappa}_\lambda$. Therefore, these three conditions
together ensure GSI occur in sequence. The detail proof is given
in Appendix B.

\begin{figure*}
\includegraphics{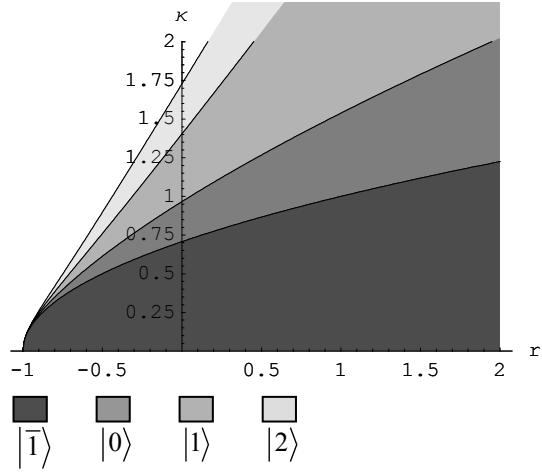}% Here is how to import EPS art
\caption{\label{fig:epsart} Regions of ground states with
different $r$. The different shadow regions correspond to
different ground states. For example, the darkest region denotes
the system ground state as $|\bar{1}\rangle$ ect. The lowest curve
is $\kappa_{\bar{1}}=\sqrt{\frac{1+r}{2}}$ obtained by solving
$E_0=-1$. }
\end{figure*}

It is interesting to point out that by adjusting the detuning
parameter $r$, it is possible to have GSI in the small coupling
regime. We recall the fact that, for on-resonant case($r=0$), the
first ground state transition occurs at
$\kappa_{\bar{1}}=\sqrt{1/2}$. However, with negative
$r$($\omega<\omega_0$), the critical
$\kappa_{\bar{1}}=\sqrt{(1+r)/2}$ is smaller than $\sqrt{1/2}$
resulting with GSI at weak coupling. On the contrary, if $\omega$
is larger than $\omega_0$, then one needs a strong coupling
strength to obtain GSI. Let $\kappa_i^<$, $\kappa_i^0$ and
$\kappa_i^>$ be the $i$th GSI critical couplings for $-1<r<0$,
$r=0$ and $r>0$ respectively. The above discussion on
$\kappa_{\bar{1}}$ can also be extended to all other cases. One
can deduce that for $-1<r<0$, $\kappa_i^<<\kappa_i^0$, while for
$r>0$, $\kappa_i^>>\kappa_i^0$. These results are numerically
shown in Fig.(1) and the proof of these general results is the
content of Appendix C. By introducing detuned frequency, one might
control the system ground state entanglement by GSI as shown in
the next section.

\section{GSI v.s. Entanglement in DSBM: Two-spin Case}

\begin{figure*}
\includegraphics{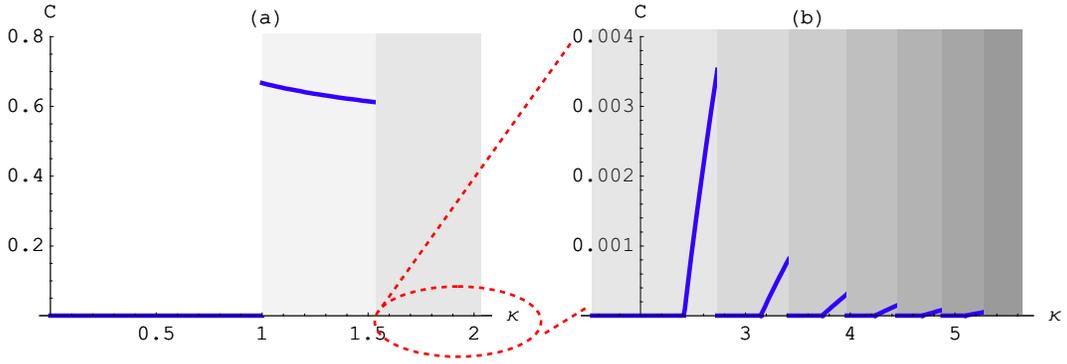}% Here is how to import EPS art
\caption{\label{fig:wide} The concurrence with $r=1$. Different
shading regions correspond to different GSI. The maximum of
entanglement appears in the interval
$[\kappa_{\bar{1}},\kappa_0]$. However, for other GSI regions, the
concurrence are strongly suppressed.}
\end{figure*}

Due to the fact that spins are coupled to the boson field, such
interaction induces quantum correlation among spins. Hence, the
spin system (by tracing out all boson states) is in general
entangled. For $2\times 2$ bipartite system, to quantify
entanglement, it has been proposed by Wootters\cite{woo}, by using
the concurrence of the system density matrix $\rho$, which is
defined by $C(\rho)=Max\{0,\xi_1- \xi_2-\xi_3-\xi_4 \}$ where
$\xi$'s are the square root of the eigenvalues of spin flow matrix
$R$ defined by $\rho$: $R=\rho(\s_y^A\otimes\s_y^B)
\rho^*(\s_y^A\otimes\s_y^B)$, subtracting in decreasing order. One
should note that, after partially tracing out the boson degree of
freedom, the spin density matrix belongs to the class of the
generalized Werner state\cite{Werner} which is defined as \bea
\rho_A = Tr_p\{|j,\lambda,h\rangle\langle j,\lambda,h|\}=
\left(\begin{array}{cccc}
\rho_{11}&0&0&0\\0&\rho_{22}&\rho_{23}&0\\0&\rho_{32}&\rho_{33}&0\\
0&0&0&\rho_{44}\end{array}\right). \nn \eea Due to the
superposition of the triplet state with $j=1$, one has
$\rho_{22}=\rho_{33}=\rho_{23}=\rho_{32}$. This density matrix has
a simple formula for the concurrence
$C(\rho_A)=2Max\{\rho_{22}-\sqrt{\rho_{11}\rho_{44}},0\}$ which is
determined by the competition of the populations between the
entangled triplet and unentangled states. The form of this density
matrix is invariant under time evolution\cite{Tanas} of $H$ given
in Eq.(1). For $\kappa_{\bar{1}}\leq \kappa<\kappa_0$, the
concurrence is \bea C_0= \frac{(r+\sqrt{8\kappa^2+r^2}
)^2}{2\{8\kappa^2+ r(r+\sqrt{8\kappa^2+r^2})\}}. \eea  For
$\kappa_{\lambda-1}\leq \kappa<\kappa_{\lambda}$ with $\lambda\geq
1$,
\begin{widetext} \bea
C_{\lambda}=\frac{3\sqrt{\lambda}\kappa^2\{3(4\sqrt{(1+\lambda)^3}
\kappa^2+\sqrt{\lambda}r^2)-4
\sqrt{3\a_{\lambda}}(\sqrt{1+\lambda}-\sqrt{\lambda})r
\zeta_{\lambda}-4\a_{\lambda}(2\sqrt{1+\lambda}-
\sqrt{\lambda})\zeta^2_{\lambda}\}}{9\kappa^2
(2(1+2\lambda)(1+\lambda)\kappa^2+\lambda
r^2)-12\sqrt{3\a_{\lambda}}\kappa^2r\zeta_{\lambda}
-6\a_{\lambda}(2(2+\lambda)\kappa^2-r^2)\zeta^2_{\lambda}+
8\a_{\lambda}^{\frac{3}{2}}\zeta^3_{\lambda}(\sqrt{3}r+
\sqrt{\a_{\lambda}}\zeta_{\lambda})} \nn \\ \eea
\end{widetext} where $\zeta_{\lambda}\equiv \cos\{
\frac{\pi}{3}-\frac{\varphi_\lambda}{3}\}$. One can easily check
that, when $r=0$, the concurrences become \bea
C_{\lambda}(r=0)=\frac{(\sqrt{1+\lambda}-\sqrt{\lambda})^2}{2(1+2\lambda)}
\eea which are positive and non-vanishing for all $\lambda$.
Therefore, ground states for \textit{all} $\kappa$ with $r=0$ are
entangled. However, it will be shown in below that this is not
true for $r\neq 0$. Note that, for $r=0$, the concurrence indeed
has discontinuity whenever there is a ground state
transition(quantum phase-like transition)\cite{Buzek} for systems
of \textit{finite number} of spins(except for $N=1$). Furthermore,
in between GSI the concurrence is a constant, for example the
concurrence for $\kappa_{\bar{1}}<\kappa<\kappa_0$ is
$\frac{1}{2}$. The $\kappa$-independence is due to the fact that
the energy eigenstates is $\kappa$-independent for $r=0$. However,
if two spins couple with an off-resonant mode, the characteristic
of the concurrence is different from the on-resonant case. Since
the eigenstates become $\kappa$-dependent, the concurrence is an
explicit function of $\kappa$. Indeed this is clearly shown in
Fig.(2) where the concurrence of the $r=1$ case is plotted. In
addition, it can be seen that for
$\kappa_{\bar{1}}<\kappa<\kappa_0$, one has $C_0>0.5$ which
indicates that the entanglement between spins are more enhanced
than the on-resonant result. It is also shown in Fig.(2) that the
concurrence is strongly suppressed for $C_{\lambda\geq 1}$. This
is a general tendency which also holds for $r=0$. Moreover, as
shown in Table I, $C_0$ becomes larger as $r$ increases. The
suppression of $C_{\lambda\geq 1}$ and the enhancement of $C_0$
can be understood by considering the ground state eigenvectors. Up
to a normalization constant, states which might become the ground
state can be expressed as follows(we neglect the
labelling of $j,h$ in the eigenkets): \bea |0\rangle &=& a_0|0\rangle |0\rangle_p+|\bar{1}\rangle|1\rangle_p \\
a_0 &=& \frac{r+\sqrt{8\kappa^2+r^2}}{2\sqrt{2}\kappa} \nn
\\ |\lambda\rangle &=& a_{\lambda}|1\rangle|\lambda-1\rangle_p+b_{\lambda}|0\rangle|\lambda\rangle_p
+|\bar{1}\rangle|\lambda+1\rangle_p \nn \\ && \\ a_{\lambda} &=&
-\frac{\sqrt{1+\lambda}}{\sqrt{\lambda}}+\frac{(\frac{2}{3}\sqrt{3\a_{\lambda}}
\zeta_{\lambda})(r+\frac{2}{3}\sqrt{3\a_{\lambda}}\zeta_{\lambda})}{2\sqrt{\lambda(1+\lambda)}\kappa^2}
\nn \\ b_{\lambda} &=& \frac{1}{\sqrt{2(1+\lambda)}\kappa }
(r+\frac{2}{3}\sqrt{3\a_{\lambda}}\zeta_{\lambda}) \nn \eea where
$\lambda\geq 1$. The expression of $a_0$ in Eq.(9) indicates the
entangled triplet state has a large amplitude as $r$ increases and
as a result enhanced entanglement arises for $C_0$. This is due to
the fact that it needs a stronger coupling strength to achieve GSI
for large detuning(See Fig.(1)), therefore, stronger
correlation(entanglement) exists. Note that this fact is also
consistent with the results of Appendix C which requires, for
large $r$, strong coupling constant for the occurrence of GSI. The
eigenkets for $\lambda\geq 1$(See Eq.(10)) are different from the
$\lambda=0$ state(Eq.(9)) by having an extra term, namely
$|1\rangle|\lambda-1\rangle_p$. Consequently, the existence of
such term is the source of diluting the entanglement of the
system. In passing, observe that, in Eq.(7) the numerator vanishes
by cancellation in the large $\lambda$ limit. It can also be
understood by noting that, at large $\lambda$, the resulting
eigenket becomes: \bea |\lambda\rightarrow\infty\rangle &=&
\{|1\rangle|+\sqrt{2}|0\rangle+ |\bar{1}\rangle\}
|\lambda\rangle_p. \nn \eea Therefore, the spin state inside the
curly bracket is a separable state which implies $C=0$.

\begin{table} \caption{\label{tab:table1}The $r$-dependence of
concurrence for $N=2$.}
\begin{ruledtabular}
\begin{tabular}{rrrr}
r&$C_0$\footnotemark[1]&$C_1$&$C_2$\\
\hline
-0.9 & 0.0977& 0.0425 & 0.0327 \\
-0.5 & 0.3613 & 0.0626 & 0.0273 \\
0    & 0.5      & 0.0286 & 0.0101 \\
0.5  & 0.5691 & 0.0124 & 0.0040 \\
1    & 0.6667 & 0.0035 & 0.0008  \\
1.2  & 0.6875   & 0.0010 & 0  \\
1.3  & 0.6970  & 0 & 0  \\
\end{tabular}
\end{ruledtabular}
\footnotetext[1]{Generally, $C_i$ is $\kappa$-dependent, the data
here are only showing the maximum values for each GSI region. This
footnote also applies to the other tables when it is appropriate.}
\end{table}
\begin{table}
\caption{\label{tab:table1}The $r$-dependence of concurrence for
$N=3$.}
\begin{ruledtabular}
\begin{tabular}{rrrrrrr}
r&$C_0$&$C_1$&$C_2$&$C_3$&$C_4$ \\
\hline
 6   & 0.5833 & 0.2944 & 0.0029 & 0.0007 & 0.0002  \\
 7.2 & 0.5942 & 0.3126 & 0.0017 & 0.0002 & 0       \\
 8   & 0.6    & 0.3233 & 0.0011 & 0      & 0        \\
10   & 0.6111 & 0.3460 & 0      & 0      & 0        \\
\end{tabular}
\end{ruledtabular}
\end{table}

One important point should be addressed is the connection between
GSI and concurrence. It has recently been discussed in the
literatures whether concurrence is a good measure to quantify
QPT\cite{Mosseri,Yang,Milburn,Buzek,Martin}. For the case with
$r=0$, it has been shown that the discontinuity of concurrence is
indeed associated with GSI even for \textit{finite}
system\cite{Buzek}. However, by analyzing the $r=1$ case
carefully, one can see from Fig.(2) that there are extra
discontinuities(See Fig.(2b)) which do not relate to GSI at all.
They appear at the values of $\kappa$ where the derivative of
concurrence is discontinuous. Such non-analytical behavior is due
to the requirement that the concurrence is non-negative, but not
from non-analyticity of the density matrix. In fact this
phenomenon has also been pointed out by Mosseri et
al\cite{Mosseri} in the model of spins interacting on a simplex
embedding in a magnetic field and by Yang\cite{Yang} for the case
of XXZ chain. Furthermore, Verstraete, Martin-Delgado and Cirac
have also shown recently that, in gapped quantum spin system, the
entanglement length is diverging without quantum phase
transition\cite{Martin}. These results indicate that GSI and
concurrence are not necessary in concord with one another.
Surprisingly, a new interesting evidence is also obtained in DSBM.
To clarify further on the relation between GSI and concurrence,
the results of $C_{\lambda}$ for $\lambda=0,1$ and $2$ with
different $r$ are given in Table I. From the table, it is noted
that as $r=1.2$ the concurrence of all $\lambda\geq 2$ states
vanishes. As $r$ increases further, one can see more states
possess vanishing concurrence(For example, $r=1.3$ in Table I).
Obviously, this is different from the above conclusion which shows
the uncorrelated non-analyticity of $C$ with GSI. Here,
\textit{the analyticity of concurrence is guaranteed by $C=0$,
however, the system still shows GSI as $\kappa$ varies}. In order
to check this conclusion is not just the artifact of $N=2$ system,
the results of $N=3$ are reported in Table II, where pair-wise
concurrences\cite{Wang} have been calculated. Similar to $N=2$,
for $N=3$, one can see that $C_0$ and $C_1$ are $r$ increasing
function. Moreover, the ground states of higher GSI with $C=0$
also appear. This can be seen from Table II, as $r=7.2$, the
concurrence $C_i=0$, $i\geq 4$. With further increasing $r$, more
vanishing $C_i$ appear. This might be a finite system effect. Due
to the limited spin space, after the $N$th transition, all ground
states do not change qualitatively and hence the ability of
creating entanglement is restricted even in the strong coupling
regime. Here, we emphasize that the 1-1 correspondence between GSI
and the discontinuities of concurrence seems just a special result
for finite $N$ SBM(or DM) and can not be extended in DSBM.
Consequently, these results clearly establish the fact that 1-1
correspondence between GSI and concurrence can not be true in
general for finite $N$ DSBM. It is certainly interesting to see if
the 1-1 correspondence remains valid as $N\rightarrow\infty$.

\begin{figure*}
\includegraphics{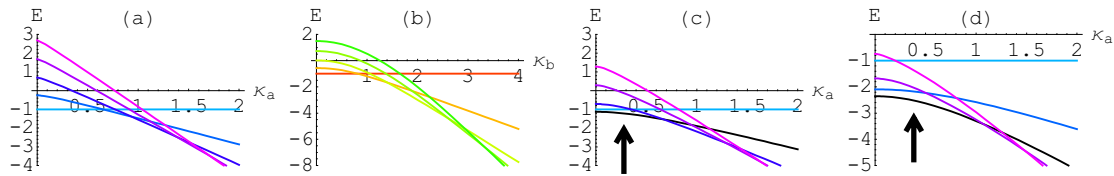}% Here is how to import EPS art
\caption{\label{fig:wide} The eigenenergy of $\lambda=-1,0,1,2,3$
for $r=1$. (a) $\kappa_b=0.4$; (b) $\kappa_a=0.4$; (c)
$\kappa_b=1.1$; (d) $\kappa_b=1.8$. The black arrows in $(c)$ and
$(d)$ indicate the eigenstate with $\lambda=0$ and $\lambda=1$
respectively.}
\end{figure*}

\section{GSI with 1+1 Modes}

\begin{table} \caption{\label{tab:table1} The critical coupling constants
for $\kappa_a=\kappa_b$ in the 1+1  spin-boson model. }
\begin{ruledtabular}
\begin{tabular}{rrrrr}
$r$&$\kappa_{\bar{1}}$&$\kappa_0$&$\kappa_1$&$\kappa_2$\\
\hline
 0\footnotemark[1]   & 0.7071 & 0.9660 & 1.4029 & 1.7260 \\
-0.9 & 0.2132 & 0.2248 & 0.2371 & 0.2498 \\
-0.1 & 0.4867 & 0.6586 & 0.9425 & 1.1569 \\
 0.1 & 0.5118 & 0.7043 & 1.0354 & 1.2758 \\
 1   & 0.5774 & 0.8158 & 1.2518 & 1.5477  \\
10   & 0.6770 & 0.9393 & 1.3910 & 1.7197 \\
100  & 0.7036 & 0.9630 & 1.4012 & 1.7247  \\
\end{tabular}
\end{ruledtabular}
\footnotemark[1] The critical $\{\kappa_i\}$ of the single mode
model with $r=0$.
\end{table}

In the previous sections, ground state instabilities of DSBM and
its correlation with concurrence have been treated. It is
interesting to see if the conclusions still hold for the case of
multi-mode spin-boson model. In this section the case of two-mode
(1+1 mode) model will be analyze by adding one off-resonant mode
to the resonant SBM. The Hamiltonian is \bea H&=&H_0+H_I
\nn \\ H_0&=&J_z+a^{\dagger}a+b^{\dagger}b\nn \\
H_I &=& rb^{\dagger}b+ \kappa_aJ_+a+\kappa^*_aJ_-a^{\dagger} \nn \\
&& +\kappa_bJ_+b+\kappa^*_bJ_-b^{\dagger} \nn \eea where
$\{b,b^{\dagger}\}$ are annihilation and creation boson operator
of the off-resonant mode with $\kappa_b$ being the coupling
constant. Similar to the single mode model, it is easy to check
$[H_0,H_I]=0$ and one can choose the eigenstates of $H_0$ to
represent operator $H_I$: \bea H_0|j,m\rangle|n_a,n_b\rangle_p
=\lambda|j,m\rangle|n_a,n_b\rangle_p \nn \\
H_I|\lambda,h\rangle=h|\lambda,h\rangle \nn \\
H|\lambda,h\rangle=E_{\lambda h}|\lambda,h\rangle \nn \eea where
$\lambda=n_a+n_b+m$ and $E_{\lambda h}=\lambda+h$. The dimension
of $H_I$ is 1 for $\lambda=-1$ and $3(\lambda+1)$ for $\lambda\geq
0$. Similarly, we will omit the labelling, $j$ and $h$, of the
eigenkets in the following discussions. GSI of the 1+1 model with
$r=1$ is shown in Figs.(3). For example, by fixing $\kappa_b=0.4$
the sequential GSI, namely $\{E_{\bar{1}}\rightarrow
E_0\rightarrow E_1\rightarrow E_2\rightarrow\ldots\}$ are clearly
shown in Fig.(3a) ( The result of keeping $\kappa_a$ fixed at 0.4
is shown in Fig. (3b)). However, as $\kappa_b$ increases, the
state $|\bar{1}\rangle= |\bar{1}\rangle|0_a,0_b\rangle_p$ may no
longer be the ground state. Therefore the sequence of GSI does not
have to begin from $\lambda=-1$. This result can be seen in
Fig.(3c) which shows the sequence of GSI with $\kappa_b=1.1$ as
$\{E_0\rightarrow E_1\rightarrow E_2\rightarrow\ldots\}$. This
fact just reflects the result of GSI in DSBM discussed in Sec.III.
Moreover, if $\kappa_b$ is further increased, more lower spin
sector eigenstates get kicked out of the GSI sequence, this is
shown in Fig.(3d) with $\kappa_b=1.8$.

It is also interesting to see how the detuning parameter is
related to the pattern of GSI. By adding one extra off-resonant
mode to SBM, the results are shown in Table III which also
includes the results of SBM for comparison. To keep things simple,
the results are evaluated with $\kappa_a=\kappa_b$ without losing
generality. One can see that the critical values $(\kappa_i)$ for
GSI are increasing function of $r$ but being bounded by the
results of SBM(See the first line of the table). Similarly, for
$r$ closed to $-1$, $\kappa_i$ also approach the results of SBM
which can be determined explicitly from Fig.(1). These interesting
results can also be understood from the energy spectrum. For
example, consider $\kappa_{\bar{1}}$ which indicates the crossing
of the eigenenergies of $|\bar{1}\rangle$ and $|0\rangle$. When
$r$ is large, the off-resonant mode has higher energy and its
excitation costs more energy. Thus, $|0\rangle$ involves
dominantly the lower energy boson which is the on-resonant mode.
As a consequence, the determination of $\kappa_{\bar{1}}$ is
governed by SBM. Physically, what is happening is the effect that
is well-known in most physical systems, namely, the result of
decoupling effect of far off-resonant driving. On the other hand,
at the limit of $r\rightarrow -1$, the off-resonant mode with
lower frequency is dominating. Therefore, $|0\rangle$ can have
more contributions from the off-resonant photon and consequently
the determination of $\kappa_{\bar{1}}$ is dictated by DSBM. It is
important to point out that with extra mode, \textit{$\kappa_i$
can be reduced significantly}. For example, as $r=1$,
$\kappa_{\bar{1}}=0.5774$ which is smaller than $0.7071$ and $1$
of the critical couplings $\kappa_{\bar{1}}$ with $r=0$ and $r=1$
respectively(See Fig.(1)). Since for $\kappa>\kappa_{\bar{1}}$ the
ground state is entangled, it is certainly important to obtain GSI
in the weak coupling regime. However, having GSI at lower critical
coupling is not enough for practical reasons. One important
requirement for employing entanglement in quantum information
science is to have strong enough entanglement or maximally
entangled state. Hence, it is necessary to see if adding extra
mode can either enhance or suppress entanglement.

\begin{table} \caption{\label{tab:table1}The $r$-dependence of concurrence
in 1+1 spin-boson model.}
\begin{ruledtabular}
\begin{tabular}{rrrr}
$r$&$C_0$&$C_1$&$C_2$ \\
\hline
0    & 0.5    & 0.0286 & 0.0101  \\
-0.9 & 0.1074 & 0.0508 & 0.0394  \\
-0.1 & 0.4898 & 0.0324 & 0.0114  \\
 1   & 0.5455 & 0.0211 & 0.0075  \\
 1.8 & 0.5462 & 0.0226 & 0.0084  \\
 5   & 0.5381 & 0.0316 & 0.0135  \\
10   & 0.5253 & 0.0348 & 0.0161  \\
\end{tabular}
\end{ruledtabular}
\end{table}

The $r$-dependent results of few $C_i$'s are tabulated in Table IV
which also contains the results of SBM(the first row). For
practical aspect, we only concentrate on $C_0$ which has higher
entanglement. One can see that the maximum value is around
$r=1.8$. However, at larger $r$, $C_0$ is decreasing toward the
value of the SBM result. This result is the same decoupling effect
discussed previously and once again the on-resonant mode
determines $C_0$. Note that the maximal value of $C_0$ is higher
than the one in SBM and the determination of the value of $r$ with
the maximal concurrence is a balanced result of the competition
between two modes. The same effect happens for other $C_i$ in
Table IV. One should note that the entanglement obtained in the
weak coupling region is distillable\cite{Distillable} and can be
\textit{enhanced} comparing with the on-resonant result($C=0.5$).
Therefore, one can have an "entanglement switch"\cite{scaling} by
controlling the first ground state transition and it seems that
\textit{having extra mode can do just that}. All in all, this
result seems to suggest that 1+1 mode might be functioning better
than mono-mode models. It is important to justify the above
conjecture by studying a system with three cavity modes which will
be reported elsewhere.

\begin{table} \caption{\label{tab:table1} The results of 1+1 cavity modes
with $r_a=1.2$ and $\kappa_a=\kappa_b$.}
\begin{ruledtabular}
\begin{tabular}{rrrrr}
 $r_b$&$C_0$&$C_1$&$C_2$ \\
\hline
1       & 0.6764 & 0.0023 & 0.0004 \\
1.1     & 0.6823 & 0.0016 & 0.0002 \\
1.3     & 0.6921 & 0.0005 & 0      \\
1.5     & 0.6998 & 0      & 0      \\
100     & 0.6920 & 0.0043 & 0.0035 \\
10000   & 0.6875 & 0.0011 & 0     \\
\end{tabular}
\end{ruledtabular}
\end{table}

Finally, for comparison, the results of both cavity modes being
off-resonant with $r_a=1.2$ and $r_b$ are given in Table V. The
result is quite interesting since the value of $C_0$ can be
\textit{higher} than the corresponding results($r=1.2$ with
$C_0=0.6875$) of SBM in Table I. For example, when $r_b=1.5$ one
has $C_0=0.6998$ which is larger than 0.6875 given in Table I.
Furthermore, as $r_b$ becomes very large, one can see from this
table that $C_0$ approach the value in SBM. Again this is just the
effect of decoupling. Further support of this fact is revealed the
values of $C_1$ and $C_2$ in Table V. Hence, for enhancing
entanglement, it is likely that off-resonant multi-mode model is a
better candidate and deserve further analysis. Similarly, the
absence of discontinuity of the concurrence with GSI can also be
found in the multi-mode system (See Table V). Moreover, we find
that the extra discontinuities of concurrence also appear in two
off-resonant modes of spin-boson model. These facts show more
supports for disconnecting GSI and discontinuities of concurrence
in spin-boson model.

\section{Conclusion}
In this work, we show explicitly that GSI and entanglement are not
necessarily connected. This is shown by a rigorous analysis of the
spin-boson model(SBM) with two spins. By knowing the full
spectrum, it is shown that the sequential quantum phase
transitions occur in this system and the closed form expression of
concurrence is obtained. Employing these results we are able to
clarify the the relation between GSI and entanglement. Contrary to
the results in literatures which are concentrated on spin-spin
interating systems, it is shown clearly that in the detuned
spin-boson model, concurrence is not a good measure for
quantifying GSI. This is shown by realizing that \textit{not all}
the discontinuity of concurrence are associated with the ones
appearing in GSI and on the contrary the system having GSI can be
corresponding to a continuous concurrence. Even though the above
results are obtained for $N=2$ system, we have also obtained
numerical results for $N=3$ which also support our conclusion.
Furthermore, the 1+1 mode model is analyzed and interesting
results are obtained. It is seen that the effects of extra mode
are two folds. First of all, GSI can happen at weak critical
couplings which are important for having entangled ground state.
Secondly, the entangled ground state with extra off-resonant mode
have a higher concurrence comparing to the results of SBM and
DSBM. These results may be useful in the context of quantum
information science. Moreover, we also confirmed that GSI and the
discontinuities of concurrence remain uncorrelated even extra mode
is included. There are several directions for further study along
this work. It is interesting to obtain results for adding more
modes to justify the effects of extra modes as obtained here. It
is also necessary to analyzed the $N$ atoms case. Furthermore,
results for more than two-level system are very important. For
example, for 3-level systems, it is interesting to analyze either
the $\Lambda$ system or V system to see if the results obtained in
this work remain valid, since these systems are also quite common
in atomic physics. These problems will be pursuit in the future.

\vskip 1.0cm This work was supported by the National Science
Council of R.O.C. under the Grant No. NSC 93-2112-M-006-006 and
94-2112-M-006-016. The support from the Center for QIS at NCKU and
the Tainan Branch of the National Center for Theoretical Sciences
of R.O.C. are acknowledged.

\appendix
\section{}
\indent In this appendix, we provide the finite matrices of $H_I$
for given $\lambda$ with $j=\frac{N}{2}$. For
$\lambda<\frac{N}{2}$, the general matrix form of $H_I$ for
arbitrary $N$-atom system is: (with excitation number
$\lambda=-\frac{N}{2}+\nu$ and $\nu<N$)
\begin{widetext} \bea \left(\begin{array}{ccccccc}
0&\kappa^*\sqrt{R_{\nu}}&0&\ldots&\ldots &\ldots
&\ldots\\\kappa\sqrt{R_{\nu}}&r&
\kappa^*\sqrt{2}\sqrt{R_{{\nu}-1}}&0&\ldots &\ldots &\ldots\\
0&\kappa\sqrt{2}\sqrt{R_{{\nu}-1}}&2r &
\kappa^*\sqrt{3}\sqrt{R_{{\nu}-2}}&0 &\ldots&\ldots\\
0&\vdots&\vdots&\ddots&\ldots&\ldots &\ldots\\
0&\vdots&\vdots&\kappa\sqrt{k}\sqrt{R_{{\nu}-k-1}}&kr&
\kappa^*\sqrt{k+1}\sqrt{R_{{\nu}-k}}&\ldots \\
\vdots&\vdots&\vdots&\vdots&\vdots&\ddots &\vdots\\
0&0&0&0&\kappa\sqrt{{\nu-1}}\sqrt{R_2}&{(\nu-1)}r&\kappa^*\sqrt{{\nu}}\sqrt{R_1}\\
0&0&0&0&0&\kappa\sqrt{{\nu}}\sqrt{R_1}&{\nu}r
\end{array}\right) \nn \\ \eea where $R_x\equiv x(N+1-x)$ and $R_{x\leq 0}=0$. The
non-vanishing off-diagonal elements only exist next to the
diagonal on each row. For $\lambda\geq\frac{N}{2}$, the dimension
of the matrix is fixed which is $(N+1)\times(N+1)$. The general
matrix form of $H_I$ is \bea \left(\begin{array}{ccccc} \nu
r&\kappa^*\sqrt{\nu+1}\sqrt{R_N}&0&\ldots
\\\kappa\sqrt{\nu+1}\sqrt{R_N}&(\nu+1)r&
\kappa^*\sqrt{\nu+2}\sqrt{R_{N-1}}&\ldots\\
\vdots&\vdots&\ddots&\ldots \\
0&\ldots&\kappa\sqrt{\nu+N+1}\sqrt{R_1}&(\nu+N)r
\end{array}\right) \nn \\ \eea \end{widetext} where $\lambda=\frac{N}{2}+\nu$ and $\nu\geq 0$.
For the first ground state transition, the critical coupling for
arbitrary $N$ and detuning $r$ is \bea
\kappa_{\bar{\frac{N}{2}}}=\sqrt{\frac{1+r}{N}} \eea which can be
obtained by solving the equation of $E_{\bar{\frac{N}{2}}}=
E_{\bar{\frac{N}{2}}+1}$
where \begin{subequations} \bea E_{\bar{\frac{N}{2}}}&=& -\frac{N}{2} \\
E_{\bar{\frac{N}{2}}+1} &=& -\frac{N}{2}+1+
\frac{1}{2}\{r-\sqrt{4N|\kappa|^2+r^2}\}. \eea \end{subequations}

\section{}
\indent This appendix is to prove that not only the existence of
GSI of two-atom system with arbitrary detuning but also these GSI
occur in sequence. In order to prove these results, it is
necessary to determine the ranges of $\varphi_\lambda$ and
$\theta_\lambda$. Denoting $x\equiv(\frac{\kappa}{r})^2$ and
recalling the definition of $\varphi_\lambda$, one has: \bea
\varphi_\lambda&\equiv&\cos^{-1}\{\frac{3\sqrt{3}\kappa^2
r}{\sqrt{\a_\lambda^3}} \} \nn\\ &=&\cos^{-1}\{
\frac{\sgn(r)3\sqrt{3}x}{\sqrt{1+6\tau_{\lambda}x+12\tau_{\lambda}^2x^2
+8\tau_{\lambda}^3x^3}}\} \eea where $\tau_{\lambda}\equiv
1+2\lambda$ and $\sgn(r)$ is the sign function. Since
$0\leq\kappa<\infty$ and $-1<r<\infty$, one has $0\leq x<\infty$.
The ranges of $\varphi_{\lambda}$ and $\theta_{\lambda}$ are: \bea
\cos^{-1}(\frac{1}{\tau_{\lambda}}) \leq\varphi_{\lambda}\leq
\cos^{-1}(\frac{-1}{\tau_{\lambda}}) \\ \theta_\lambda^-\leq
\theta_{\lambda}\leq\theta_{\lambda}^+\\\theta_{\lambda}^{\pm}
\equiv\frac{1}{3} \{\pi-\cos^{-1}(\frac{\pm 1}{\tau_{\lambda}})\}.
\eea Although the maximal value of $\theta_{\lambda}$ is $\lambda$
dependent, it is easy to check that $\theta_{\lambda}$ is bounded
as follows: \bea\theta_1^-\leq\theta_{\lambda}\leq\theta_1^+.\eea
By knowing the ranges of the angles, we are now in the position of
showing the conditions $(i)$, $(ii)$ and $(iii)$.

For $(i)$, it is easy to see that $E_0$ is a monotonic decreasing
function by directly checking $\partial_{\kappa}E_0$:
\bea\partial_{\kappa}E_0= -\frac{4\kappa}{\sqrt{8\kappa^2+r^2}}
\eea which is negative definite for $\kappa>0$ and $-1<r<\infty$.
For $E_{\lambda}$ given by Eq.(5c) with $\lambda\geq1$, the
partial $\kappa$-derivative of $E_{\lambda}$ is: \bea
\partial_{\kappa}E_{\lambda} &=& -\Omega_{\lambda}\chi_{\lambda} \\\Omega_{\lambda}
&\equiv&\frac{4\kappa}{\sqrt{3\a_{\lambda}}\sqrt{
\a_{\lambda}^3-27\kappa^4r^2}}, \nn \\ \chi_{\lambda} &\equiv&
\tau_{\lambda}\zeta_{\lambda}\sqrt{\a_{\lambda}^3-27\kappa^4r^2}
+\sqrt{3}r(\tau_{\lambda}\kappa^2-r^2)\sin\theta_{\lambda}.\nn
\eea Note that, $\a_{\lambda}^3>27\kappa^4r^2$ and
$\a_{\lambda}\geq 0$, then $\Omega_{\lambda}\geq 0$ for all
$\kappa$ and $r$. Moreover, by using the ranges of the angles, one
has \bea
0.804\sim\cos\theta_1^+\leq\cos\theta_{\lambda}\leq\cos\theta_1^-\sim
0.917 \nn \\0.399\sim\sin\theta_1^-\leq\sin\theta_{\lambda}
\leq\sin\theta_1^+\sim 0.595 \nn\eea Therefore,
$\cos\theta_{\lambda}>\sin\theta_{\lambda}>0$. For $r\geq 0$ and
$\tau_{\lambda}\geq 3$, one has $\chi_{\lambda}>\chi_A-\chi_B$
with \bea \chi_A &=&
\sqrt{3(\a^3_{\lambda}-27\kappa^4r^2)\cos^2\theta_{\lambda}}+3\kappa^2r\sin\theta_{\lambda}
\nn \\ \chi_B &=& \sqrt{3r^6\sin^2\theta_{\lambda}}, \nn \eea By
expanding out $\a_{\lambda} $ in $\chi_A$ and regrouping terms one
obtains \bea \chi_A=\sqrt{3r^6\cos^2\theta_{\lambda}
+\Delta}+3\kappa^2r\sin\theta_{\lambda} \nn \eea where $\Delta$
denotes the remaining positive part inside the square root.
Obviously, $\cos^2\theta_{\lambda}>\sin^2\theta_{\lambda}$, one
concludes $\chi_A>\chi_B \Rightarrow \chi_{\lambda}\geq 0$ for all
$\kappa$ and $r\geq 0$. Similarly, it is easy to check, for
$-1<r<0$, $\chi_{\lambda}\geq 0$ for all $\kappa$. Therefore,
$E_{\lambda}$ are monotonic decreasing function for $\lambda\geq
0$.

To show $f(\kappa,r,\lambda)=E_{\lambda+1}-E_\lambda$ is a
monotonic decreasing function of $\kappa$ for any $r$ and
$\lambda$, it is necessary to break down the proof for different
regions of $\lambda$. For the $\lambda=-1$ case, by Eq.(5),
\bea\partial_{\kappa}f(\kappa,r,\bar{1})
&=&\partial_{\kappa}(E_0-E_{\bar{1}}) \nn \\ &=&
\partial_{\kappa}E_0 \eea which is monotonic decreasing as proved in
the criterion $(i)$. One can easily check $f(\kappa,r,\bar{1})$
approaches $1$ at small $\kappa$, while becomes $-\infty$ at
larger $\kappa$. Therefore, the critical coupling
$\tilde{\kappa}_{\bar{1}}$ determined by $f(\kappa,r,\bar{1})=0$
uniquely exists. For $\lambda\geq 1$, one should show that the
$\kappa$ derivative of $f$ does not change sign for all $\kappa$.
Alternatively, it is equivalent to show the function
$g(\kappa,r,\lambda)\equiv
\partial_{\kappa}E_{\lambda}$ being a monotonic \textit{decreasing} function
in $\lambda$, such that it ensures
$g(\kappa,r,\lambda+1)-g(\kappa,r,\lambda)=\partial_{\kappa}
E_{\lambda+1}-\partial_{\kappa} E_{\lambda}=\partial_{\kappa}
f<0$. By using Eq.(B7), \bea && \partial_{\lambda}
g(\lambda,\kappa,r) =
\partial_{\lambda}\partial_{\kappa} E_{\lambda} = -\frac{4\kappa y_2^{\lambda}}{3y_1^{\lambda}
\sqrt{\a_{\lambda}}\sqrt{\a_{\lambda}^3-27\kappa^4r^2}}
\times\nn\\&&\{\sqrt{3}(
3\kappa^2r+\sqrt{y_1^{\lambda}})\cos\theta_{\lambda}-3(\kappa^2r-\sqrt{y_1^{\lambda}})\sin\theta_{\lambda}
\}\nn \eea where \bea
y_1^{\lambda}&=&8\tau_{\lambda}^3\kappa^6+3(16\lambda^2+16\lambda-5)
\kappa^4r^2+6 \tau_{\lambda}\kappa^2r^4+r^6 \nn \\
y_2^{\lambda}&=&4\tau_{\lambda}^3\kappa^6+2(16{\lambda}^2+16\lambda-5)
\kappa^4r^2+5 \tau_{\lambda}\kappa^2r^4+r^6. \nn \eea For $r\geq
0$, $\kappa^2r<\sqrt{y_1^{\lambda}}$, one has
$\partial_cg(\lambda,\kappa,r)<0$. Similarly,
$\partial_cg(\lambda,\kappa,r)<0$ is still true for $-1<r<0$ due
to $3\kappa^2|r|<\sqrt{y_1^{\lambda}}$. Thus one has
$\partial_cg<0$ for all $\kappa$, $\lambda$ and $r$. Therefore, we
have shown $g(\kappa,r,\lambda)$ is a monotonic decreasing
function of $\lambda$ and then $f(\kappa,r,\lambda)$ is a strictly
decreasing function of $\kappa$. Furthermore, one can check that,
for $\lambda\geq 1$,
\bea f(\kappa,r,\lambda)|_{\kappa\rightarrow 0} &=& 1+r>0 \nn \\
f(\kappa,r,\lambda)|_{\kappa\rightarrow\infty} &=&
\sqrt{2+4\lambda}-\sqrt{6+4\lambda}<0. \nn \eea As a result, the
crossings $\{\tilde{\kappa}_i\}(i\geq 1)$ has unique solution.

To prove the remaining case with $\lambda=0$, we express
$\partial_{\kappa}f(\kappa,r,0)$ in terms of $x$: \bea
\partial_{\kappa}
f(\kappa,r,0)&=&\partial_{\kappa} (E_1-E_0) = \eta\Ga \nn \\
\eta &\equiv& \frac{4}{\sqrt{(8x+1)(6x+1)}} \nn \\ \Ga &\equiv&
\sqrt{x(6x+1)}-\sqrt{3x(8x+1)}\cos\theta_1
\nn\\&&-\sgn(r)(3x-1)\sqrt{\frac{x(8x+1)}{\tilde{y}}}\sin\theta_1
\nn \eea where $\tilde{y}\equiv y_1^1/r^6$. If $\Ga$ is negative
for all $x\geq 0$, then $f(\kappa,r,0)$ is monotonic decreasing.
Let us start with $\sgn(r)=+$. For $x\geq\frac{1}{3}$,
$\sqrt{6x+1}<\sqrt{8x+1}$ and $\sqrt{3}\cos\theta_1>1 $, then
$\sqrt{x(6x+1)}< \sqrt{x(8x+1)}\sqrt{3} \cos\theta_1$. Therefore,
we have $\Ga<0$ for $x\geq\frac{1}{3}$and $r\geq 0$. Similarly one
has $\Ga<0$ for $x\in[(0,\frac{1}{3})$ with $\sgn(r)=-$. However,
for $x\in[0,\frac{1}{3}]$ with $\sgn(r)=+$, the last term of $\Ga$
is negative, thus it is not obvious that $\Ga$ is negative
definite. Therefore a different approach is called for. To proceed
further for $x\in[0,\frac{1}{3}]$ with $\sgn(r)=+$, let \bea
A&\equiv& \sqrt{x(6x+1)}\nn\\B&\equiv&(1-3x)
\sqrt{\frac{x(8x+1)}{\tilde{y}}}\sin\theta_1\nn\\
C&\equiv&\sqrt{3x(8x+1)}\cos\theta_1\nn, \eea and then
$\Ga=A+B-C$. Our logic to prove $\Ga$ is still negative is to show
that there exists a $\d\geq 0$ such that $\d C\geq A$ and
$(1-\d)C\geq B$, then $C\geq A+B\Rightarrow \Ga\leq 0$ for
$x\in(0,\frac{1}{3})$. To begin with, one has \bea \d C\geq A
\Rightarrow \d &\geq& \frac{A}{C} \nn \\ \d &\geq&
\sqrt{\frac{6x+1}{3(8x+1)}}\sec\theta_1
\eea The other condition is \bea (1-\d) &\geq& \frac{B}{C} \nn\\
\d &\leq& 1-\frac{1-3x}{\sqrt{3\tilde{y}}}\tan\theta_1 \eea
Combining Eq.(B9) and (B10), \bea
1-\frac{1-3x}{\sqrt{3\tilde{y}}}\tan\theta_1 \geq \d \geq
\sqrt{\frac{6x+1}{3(8x+1)}}\sec\theta_1 \geq 0\nn \eea Therefore,
$\d$ exists if the following condition is satisfied: \bea
1-\{\underbrace{\frac{1-3x}{\sqrt{3\tilde{y}}}\tan\theta_1}_{Z_1}+
\underbrace{\sqrt{\frac{6x+1}{3(8x+1)}}\sec\theta_1}_{Z_2}\}\geq
0\eea It is easy to show the ranges of $Z_1$ and $Z_2$ are: \bea
0\leq &Z_1&\leq\frac{1}{3} \nn \\ \frac{2}{3}\leq &Z_2& \leq
\sqrt{\frac{3}{11}}\sec\theta_1^+ \nn \eea Therefore, \bea
\frac{2}{3}\leq Z_1+Z_2\leq
\frac{1}{3}+\sqrt{\frac{3}{11}}\sec\theta_1^+ <1 \eea By the same
approach, one can show that $\Ga$ is negative definite for $r\geq
\frac{1}{3}$ with $\sgn(r)=-$. Furthermore, it is easy to show
that $f(\kappa,r,0)\rightarrow 1$ when  $\kappa$ is small and
changes sign at large $\kappa$. This completes the proof of
showing $\kappa_2$ exists. For now, we have shown the crossings
$\{\tilde{\kappa}_i\}$ between $E_{\lambda}$ and $E_{\lambda+1}$
exist even in the detuning two-atom system. In what following, we
will show these crossings occur in sequence.

We start from $\lambda=-1$ case, with the solution of
$E_{\bar{1}}=E_0$,
$\tilde{\kappa}_{\bar{1}}=\sqrt{\frac{1+r}{2}}$, one has: \bea &&
\tilde{E}_1-\tilde{E}_0 =
(2+r)-\frac{2}{\sqrt{3}}\sqrt{(2+r)^2-(1+r)}\cos\tilde{\theta}_1
\nn\eea where the tilde symbol denotes quantity at the appropriate
critical $\kappa$, for here it is
$\kappa=\tilde{\kappa}_{\bar{1}}$. For $\lambda=0$, imposing
$E_1=E_0$, one obtains \bea 1+r=\frac{2}{3}\sqrt{3\tilde{\a}_1}
\cos\tilde{\theta}_1+\frac{r-\sqrt{8{\tilde{\kappa}_0}^2+r^2}}{2}.
\eea And
\bea &&\tilde{E}_2-\tilde{E}_1\nn\\
&=& 1+r-\frac{2}{3}\{\sqrt{3\tilde{\a}_2}\cos\tilde{\theta}_2
-\sqrt{3\tilde{\a}_1}\cos\tilde{\theta}_1 \} \nn \\ &=&
\frac{r}{2}+\frac{4\sqrt{{6\tilde{\kappa}_0}^2+r^2}\cos\tilde{\theta}_1}{\sqrt{3}}
\{1-\frac{1}{2}\sqrt{\frac{10{\tilde{\kappa}_0}^2+r^2}{6{\tilde{\kappa}_0}^2+r^2}}
\frac{\cos\tilde{\theta}_2}{\cos\tilde{\theta}_1}
\nn\\&&-\frac{\sqrt{3}}{8}\sqrt{\frac{8{\tilde{\kappa}_0}^2+r^2}{6{\tilde{\kappa}_0}^2+r^2}}
\frac{1}{\cos\tilde{\theta}_1}\} \eea where Eq.(B13) has been used
in the second line and \bea
\cos\tilde{\theta}_1&=&\cos\{\frac{1}{3}
(\pi-\cos^{-1} \frac{3\sqrt{3}r(1+r)} {2\sqrt{(3+3r+r^2)^3}})\}\nn\\
\cos\tilde{\theta}_2&=&\cos\{\frac{1}{3}(\pi-\cos^{-1}\frac{3\sqrt{3}r(1+r)}
{2\sqrt{(5+5r+r^2)^3}})\}. \nn \eea In order to estimate the value
of Eq.(B14), By using the inequalities (B3) and (B5) one has: \bea
\frac{\cos\theta_2^+}{\cos\theta_1^-} \leq
\frac{\cos\tilde{\theta}_2}{\cos\tilde{\theta}_1} \leq
\frac{\cos\theta_2^-}{\cos\theta_1^+} \nn \\
\frac{1}{\cos\theta_1^-} \leq \frac{1}{\cos\tilde{\theta}_1} \leq
\frac{1}{\cos\theta_1^+}. \nn \eea For $\sgn(r)=+$, it is easy to
determine the minimal value of the part in the bracket: \bea
\{\ldots\}&>& 1-\frac{1}{2}\sqrt{\frac{10{\tilde{\kappa}_0}^2
+r^2}{6{\tilde{\kappa}_0}^2+r^2}}
\frac{\cos\theta_2^-}{\cos\theta_1^+} \nn\\ &&
-\frac{\sqrt{3}}{8}\sqrt{\frac{8{\tilde{\kappa}_0}^2+r^2}{6{\tilde{\kappa}_0}^2+r^2}}
\frac{1}{\cos\theta_1^+} \nn \\ &>&
1-\frac{1}{2}\sqrt{\frac{5}{3}}
\frac{\cos\theta_2^-}{\cos\theta_1^+}
-\frac{\sqrt{3}}{8}\sqrt{\frac{4}{3}} \frac{1}{\cos\theta_1^+}
\eea Then, $\tilde{E}_2>\tilde{E}_1$ at $\tilde{\kappa}_0$ for
$r\geq 0$. For $\lambda\geq 1$ cases, from
$E_{\lambda}=E_{\lambda+1}$, \bea
1+r=\frac{2}{3}\{\sqrt{3\tilde{\a}_{\lambda+1}}\cos\tilde{\theta}_{\lambda+1}
-\sqrt{3\tilde{\a}_{\lambda}}\cos\tilde{\theta}_{\lambda}\}. \eea
Therefore, \bea
&& \{E_{\lambda+2}-E_{\lambda+1}\}|_{\tilde{\kappa}_{\lambda}}\nn \\
&=&
(r+1)-\frac{2}{3}\{\sqrt{3\tilde{\a}_{\lambda+2}}\cos\tilde{\theta}_{\lambda+2}
-\sqrt{3\tilde{\a}_{\lambda+1}}\cos\tilde{\theta}_{\lambda+1}\} \nn \\
&=&
\frac{2}{3}\{2\sqrt{3\tilde{\a}_{\lambda+1}}\cos\tilde{\theta}_{\lambda+1}
-\sqrt{3\tilde{\a}_{\lambda}}\cos\tilde{\theta}_{\lambda}\nn\\&&-\sqrt{3\tilde{\a}_{{\lambda}+2}}\cos\tilde{\theta}_{\lambda+2}
\}\eea where Eq.(B14) has been used in the second line. For the
ease of discussion, we denote
$\xi(\lambda)\equiv\sqrt{3\a_{\lambda}}\cos\theta_{\lambda}$. If
the curvature of $\xi(\lambda)$ is negative, then we have
$2\xi(\lambda+1)>\xi(\lambda)+\xi(\lambda+2)$ which is just the
condition of Eq.(B15)$>0$. By taking second derivative with
$\lambda$ directly, we obtain \bea
\partial^2_{\lambda} \xi(\lambda) &=&-(\frac{\a_{\lambda}}{y_{\lambda}})^{
\frac{3}{2}}(\frac{4\kappa^4}{r^6})\times\nn\\&&\{\sqrt{3y_{\lambda}}
\cos\theta_{\lambda}+27\frac{\kappa^2}{r^2}\sin\theta_{\lambda}\}<0.\eea
Therefore, the crossing
$\{\tilde{\kappa}_{\lambda}\}|_{{\lambda}\geq 1}$ actually occur
in sequence for all $-1<r<\infty$. This complete the proof for
having sequential GSI. $\P$

\section{}
In order to prove $\kappa_i^<<\kappa_i^0<\kappa_i^>$ where
$\kappa_i^{\{<,0,>\}}$ are for $r\{<,=,>\}0$ respectively, the
hamiltonian of the system can be rearranged as: \bea
H^r=H^0+ra^{\dagger}a \eea where $H^0$ denotes the on-resonance
hamiltonian and $H^r$ is the detuned hamiltonian. $a^{\dagger}a$
is photon number operator which is positive-valued. For $r\geq 0$,
given $\kappa=\kappa_i^>$, the ground state of the system is
$|g_{i+1}^>\rangle$ such that
$H^r|g_{i+1}^>\rangle=E^r|g_{i+1}^>\rangle$. The expectation value
of $H^r$ in $|g_{i+1}^>\rangle$ can be written as: \bea E^r &=&
\langle g_{i+1}^>|H^0|g_{i+1}^>\rangle + r\langle
a^{\dagger}a\rangle_{g^>} \nn \\ &\geq& \langle g^0|H^0|g^0\rangle
|_{\kappa=\kappa_i^>}+r\langle a^{\dagger}a\rangle_{g^>} \nn \\
&\Rightarrow& E^r>E^0|_{\kappa=\kappa^>_i}. \eea Where
$|g^0\rangle$ is the ground state of $H^0$. The inequality occur
by noting that $|g_{i+1}^>\rangle$ is a trial state for $H^0$.
Moreover, due to the fact of sequential GSI proven in appendix B,
Eq.(C2) implies $\kappa_i^0<\kappa_i^>$. Similarly, for $-1<r<0$,
instead of using $|g_{i+1}^>\rangle$, we evaluate the expectation
value of $H^r$ with the ground state $|g_{i+1}^0\rangle$ of $H^0$
at $\kappa=\kappa_i^>$ and obtain the desired result : \bea E^0
&=& \langle g_{i+1}^0|H^r|g_{i+1}^0\rangle + |r| \langle
a^{\dagger}a\rangle_{g^0} \nn \\ &\geq& E^r|_{\kappa=\kappa_i^0}+
|r|\langle a^{\dagger}a\rangle_{g^0} \nn
\\ &\Rightarrow& E^r|_{\kappa=\kappa_i^0} < E^0. \eea Again the property of
trial state has been used in Eq.(C3) which implies
$\kappa_i^<<\kappa_i^0$. $\P$

%\bibliography{QPT}% Produces the bibliography via BibTeX.

\end{document}